\documentclass[prx,aps,twocolumn]{revtex4-1}

\usepackage{amsmath}  
\usepackage{amsthm}  
\usepackage{amssymb}	 
\usepackage{graphicx}  
\usepackage[usenames,dvipsnames]{color}
\usepackage{array} 
\usepackage{paralist} 
\usepackage{amsfonts}
\usepackage{mathtools}
\usepackage{times}
\usepackage{braket}
\usepackage{slashed}
\usepackage[colorlinks=true,linkcolor=blue,citecolor=blue,breaklinks]{hyperref}
\normalfont


\renewcommand{\vec}[1]{\ensuremath{\mathbf{#1}}} 
 
\newcommand{\abs}[1]{\left| #1 \right|} 
\newcommand{\avg}[1]{\left< #1 \right>} 
 
\let\baraccent=\= 
\renewcommand{\=}[1]{\stackrel{#1}{=}} 

\newcommand{\be}{\begin{equation}}
\newcommand{\ee}{\end{equation}}

\newcommand{\bea}{\begin{eqnarray}}
\newcommand{\eea}{\end{eqnarray}}
\newcommand{\beal}{\begin{align}}
\newcommand{\eeal}{\end{align}}

\newcommand{\intg}{\mathbb{Z}}

\begin{document}

\title{Emergence of a Field-Driven $U(1)$ Spin Liquid in the Kitaev Honeycomb Model}
\author{Ciar\'{a}n Hickey}
\email[E-mail: ]{chickey@thp.uni-koeln.de}
\author{Simon Trebst}
\affiliation{Institute for Theoretical Physics, University of Cologne, 50937 Cologne, Germany}

\begin{abstract}
In the field of quantum magnetism, the exactly solvable Kitaev honeycomb model serves as a paradigm for the fractionalization of spin degrees of freedom and the formation of $\intg_2$ quantum spin liquids. An intense experimental search has led to the discovery of a number of spin-orbit entangled Mott insulators that realize its characteristic bond-directional interactions and, in the presence of magnetic fields, exhibit no indications of long-range order. Here, we map out the complete phase diagram of the Kitaev model in tilted magnetic fields and report the emergence of a distinct gapless quantum spin liquid at intermediate field strengths. Analyzing a number of static, dynamical, and finite temperature quantities using numerical exact diagonalization techniques, we find strong evidence that this phase exhibits gapless fermions coupled to a massless $U(1)$ gauge field. We discuss its stability in the presence of perturbations that naturally arise in spin-orbit entangled candidate materials.
\end{abstract}
\maketitle


Quantum spin liquids are highly entangled quantum states of matter that exhibit fractionalized excitations \cite{Savary2017quantum}. A principle example for such a fractionalization are the spinon excitations of a resonating valence bond (RVB) liquid \cite{AndersonSL}, which carry spin-1/2 and arise only after breaking apart a spin-1 excitation originating from an elementary spin-flip process. Crucially, once a pair of spinons has been created in an RVB liquid, they can be separated to arbitrary distances at no energy cost -- the spinons are deconfined. 
This reveals the emergence of a much larger underlying structure present in any quantum spin liquid -- a lattice gauge theory in its deconfined regime. The interplay of fractionalization and lattice gauge theory can be conceptualized by a parton construction \cite{Wen1992}, which decomposes the original spin degrees of freedom in terms of partons that 
represent the emergent fractional degrees of freedom. These partons can be chosen to be complex Abrikosov fermions \cite{Abrikosov1965}, real Majorana fermions \cite{Tsvelik1992,Kitaev2006}, or bosons. Concomitantly, the system is found to be enriched by an emergent gauge structure, with examples including continuous $U(1)$ or discrete $\intg_2$ gauge symmetry \cite{Read1991,Senthil2000}. One of the most beautiful examples of a parton construction has been introduced by Kitaev, who was able to devise an exactly solvable spin-$1/2$ model on the honeycomb lattice with several quantum spin liquid ground states \cite{Kitaev2006}. Here, the fractionalization of the original spin degrees of freedom into Majorana fermions and an emergent $\intg_2$ gauge structure naturally appear in the framework of Kitaev's exact solution, which has led to a plethora of theoretical investigations and deep analytical insights into spin liquid physics \cite{Hermanns2018physics}.  

On a microscopic level, the key ingredients of the Kitaev model are its bond-directional Ising-type exchange interactions.
Remarkably, these seemingly unusual interactions are found to be realized via an intricate interplay of spin-orbit coupling, crystal field effects, and strong interactions \cite{Khaliullin2005orbital,Jackeli2009} in a variety of $4d$ and $5d$ materials \cite{Trebst2017}. However, these spin-orbit entangled Mott insulators are typically found to exhibit ordered states at low temperatures in lieu of the sought-after spin liquid physics, consistent with a theoretical analysis of perturbed Kitaev magnets that exhibit more conventional types of exchanges beyond a dominant bond-directional interaction \cite{Chaloupka2010,Jiang2011possible,Kimchi2011,Chaloupka2013,Rau2014,Rousochatzakis2015,Rau2016,Janssen2017}. 

Recently considerable excitement has arose due to the fact that in one of these materials, RuCl$_3$, the magnetic order can be suppressed with an in-plane magnetic field \cite{ColdeaRuClMF2015,KoichiRuClMF2015,BaenitzRuClMF2015,SearsMF2017,Hirobe2017magnetic,Kasahara2017,Banerjee2018,JanNMR2018}. 
Probably the most spectacular result is a report~\cite{KasaharaArXiv2018} for tilted field directions, which suggests that a phase, intermediate between the magnetically ordered state at low fields and the high-field polarized state, exhibits a half-quantized thermal Hall conductance -- a unique signature for a gapped topological spin liquid. The precise nature of the putative quantum spin liquid regime and its microscopic description, however, still remain open.

Motivated by these observations, we return to the original Kitaev model and explore its phase diagram in the presence of tilted magnetic fields using numerical exact diagonalization (ED) techniques. As we report in this manuscript, there are two distinct spin liquid regimes already present in this model.
For small magnetic field strengths, there is a gapped spin liquid phase whose non-Abelian topological nature has first been rationalized by Kitaev using perturbative arguments for a field pointing along the out-of-plane $[111]$ direction \cite{Kitaev2006}. Here we demonstrate that this phase is stable when tilting the magnetic field to generic directions and well beyond the perturbative regime by explicitly calculating the modular $S$-matrix from its (quasi-)degenerate ground states, which unambiguously confirms that its inherent topological nature is indeed given by the Ising topological quantum field theory (TQFT). The second spin liquid, on which we focus in this manuscript, is both manifestly distinct from the gapped topological spin liquid
and at the same time can be considered, in many ways, to be a descendent of it. As we demonstrate in this manuscript, one key distinction between the two phases is their underlying gauge structure. While the Kitaev spin liquid is accompanied by a 
$\intg_2$ gauge structure with gapped vison excitations in the gauge sector, the second spin liquid is found to exhibit the gapless gauge structure typically associated with a $U(1)$ spin liquid. By investigating the evolution of the energy spectrum, the dynamical structure factor, and thermodynamic signatures in the specific heat, we provide multi-faceted evidence that the phase transition between the two spin liquids at finite field strengths is driven by the closing of the gap for vison excitations of the $\intg_2$ spin liquid and that the emergent gapless spin liquid is a $U(1)$ spin liquid with a spinon Fermi surface. We discuss aspects of the underlying field theory governing this phase transition at the end of the manuscript. 

Finally, it should be noted that the occurrence of two stable spin liquid regimes in the Kitaev model exposed to a (tilted) magnetic field is closely linked to whether the applied field matches the underlying anti\-ferromagnetic (AFM) or ferromagnetic (FM) spin correlations, with an order of magnitude difference in the critical fields between the two cases. Only for AFM Kitaev couplings and a uniform magnetic field, do we observe the two spin liquids discussed above. For FM Kitaev couplings the gapped Kitaev spin liquid is found to be considerably less stable than in the AFM case, consistent with a number of recent numerical studies \cite{FeyMSc2013,ZhuAFMK2017,PollmannAFMK2018} (with \cite{ZhuAFMK2017} also the first to report the existence of an intermediate phase for an AFM coupling). Notably, this situation can be reversed by staggering the magnetic field, which dramatically increases the stability of the FM Kitaev phase, while the AFM spin liquid then covers a significantly smaller parameter space. To round off our discussion, we demonstrate the stability of the emergent gapless spin liquid when perturbing the Kitaev model with a conventional Heisenberg interaction or an off-diagonal $\Gamma$-exchange, which constitute further ingredients of the microscopic description of Kitaev materials \cite{Rau2014}. \\


\noindent {\bf Model --} We start our discussion by considering the pure Kitaev honeycomb model in the presence of a uniform magnetic field of arbitrary orientation, defined by the Hamiltonian
\begin{equation}
H_\pm = \pm K \sum_{\avg{i,j}\in\gamma} S_i^\gamma S_j^\gamma - \sum_i \vec{h} \cdot \vec{S}_i \,,
\end{equation}
where $H_{\pm}$ indicates an AFM/FM Kitaev coupling and the bond directions are denoted by $\gamma\in \left\{ x,y,z \right\}$. We parametrize the orientation of the magnetic field as
$\vec{h} = h\sin\theta\, \vec{\hat{h}}_{111} + h\cos\theta \,\vec{\hat{h}}_\perp$, 
where the unit vectors $\vec{\hat{h}}_{111}$ and $\vec{\hat{h}}_\perp$ point along the $[111]$ and either $[11\bar{2}]$ or $[\bar{1}10]$ directions. For materials such as (Na,Li)$_2$IrO$_3$ and RuCl$_3$ these directions correspond to the out-of plane, $c$-axis, and in-plane, $a$ or $b$-axes, respectively. The angle $\theta$ thus measures the tilt away from the honeycomb planes. \\

\begin{figure}[!t]  
\includegraphics[width=\columnwidth]{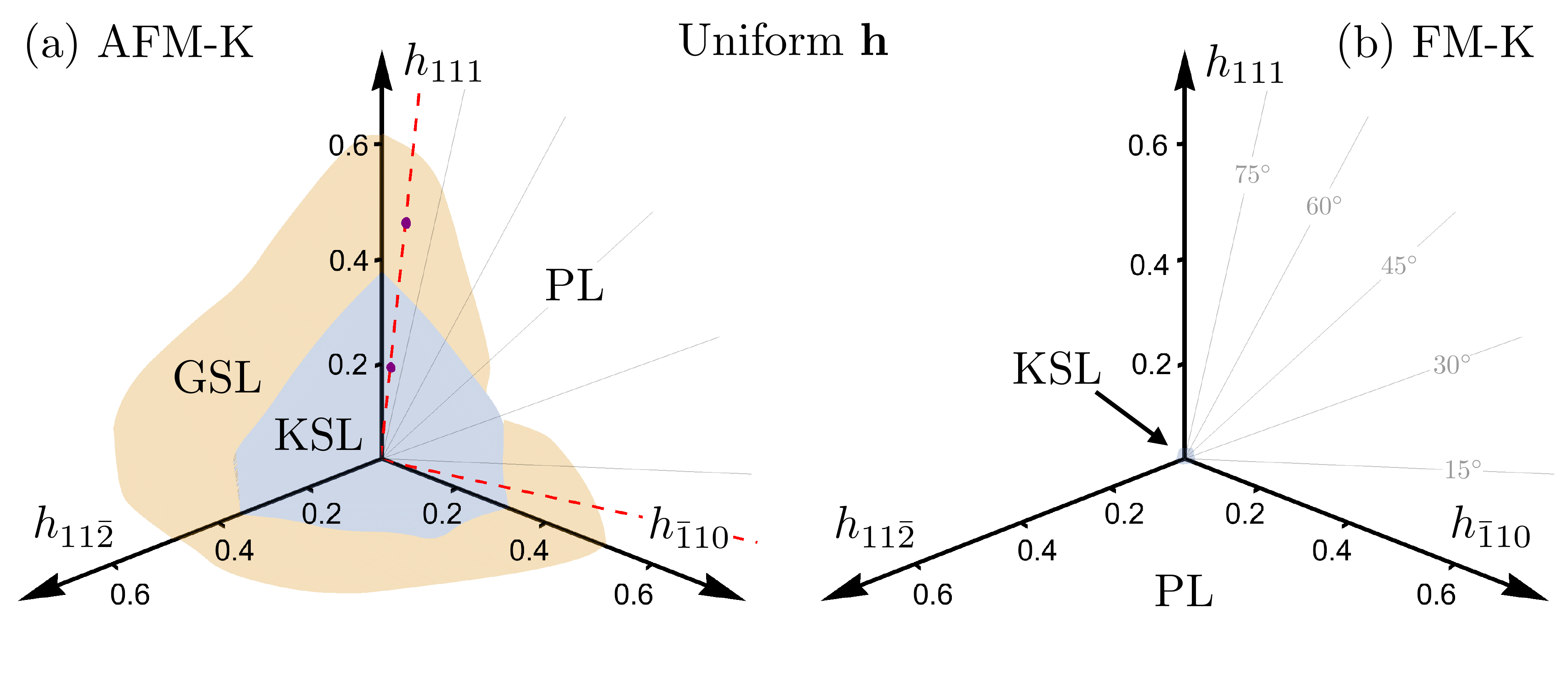}
\caption{Phase diagrams in a uniform magnetic field. (a) The pure AFM Kitaev model and (b) the pure FM Kitaev model
	     for various tilt angles. For AFM couplings the gapped Kitaev spin liquid (KSL) is surrounded, for a wide range of tilt angles, by a gapless 
	     spin liquid (GSL) before giving way to a trivial polarized state (PL). For FM couplings, in contrast, the
	     KSL is found to cover a considerably smaller parameter region with no intermediate GSL (see Supplementary Note $1$ for a zoomed-in view of the FM phase diagram). The two (purple) points in 
	    (a) mark the parameters at which the dynamical structure factor in Fig.~\ref{fig:DySF}(b) and (c) is plotted. 	  
}
\label{fig:UPDs}
\end{figure}

\noindent {\bf Phase diagrams --} The phase diagram of the model for various tilt angles of a uniform external magnetic field is presented in Fig.~\ref{fig:UPDs} for both the AFM and FM Kitaev cases. The phase boundaries, presented in this Figure, are based on a number of different signatures, including the second derivative of the ground state energy and the ground state fidelity (see Methods for more details). 
There are certain limits which have previously been discussed:\\
(i) $h=0$. In the case of zero magnetic field the Kitaev Hamiltonian is exactly solvable \cite{Kitaev2006}. Following Kitaev's original solution, each spin-$1/2$ can be split into four Majorana fermions, three are associated with the adjacent bonds
and one with  the original site. The bond Majoranas can be recombined to form a static $\intg_2$ gauge field, leaving us with a single free Majorana fermion moving in a background field. Its spectrum is gapless, with Dirac points located at the corners of the Brillouin zone, while the vison excitations of the gauge field remain gapped \cite{Kitaev2006,Hermanns2018physics}. The net result is a gapless $\intg_2$ spin liquid.

\noindent(ii) $h\parallel  [111], h \ll K $. In the presence of a magnetic field along the $[111]$ direction, Kitaev showed, using perturbation theory, that a small field opens up a gap in the Majorana spectrum. Furthermore, the resulting Majorana insulator has a non-trivial band structure, with a Chern number $C=+1$ for the lower, fully filled band. This corresponds to a gapped non-Abelian spin liquid with Ising anyon topological order, which we will refer to as the Kitaev spin liquid (KSL). The gapped flux excitations (visons) now bind a Majorana fermion and there is a single chiral gapless Majorana edge mode, which gives rise to a quantized thermal quantum Hall effect.
Our numerical data confirms that this scenario remains true away from the perturbative limit, for generic field directions, and applies to both the AFM and FM cases. Technically, we do so by calculating \cite{ZhangMES2012} the modular $S$-matrix from the three (quasi-)degenerate ground states in the KSL phase for various parameters of Fig.~\ref{fig:UPDs}. The entries $S_{ab}$ encode the braiding properties of quasiparticles $a$ and $b$ in the underlying TQFT (fixing the entries to certain universal values) and thereby allow for its unambiguous identification. Numerically, we find, e.g., the following $S$-matrix 
\begin{equation}
S_{\rm ED} = \begin{pmatrix} 0.46 & 0.74 & 0.47 \\
0.71 & 0.04 e^{-0.91 i} & -0.70 \\  
0.49 & -0.67 e^{0.02 i} & 0.58 e^{-0.13 i} 
\end{pmatrix}  ,
\end{equation}
computed for a $[111]$ field of magnitude $h \sim h^{crit}_{\rm KSL}/2$. 
For the Ising TQFT the expected $S$-matrix has corner entries $+1/2$, a middle entry of zero, and the remaining four entries $\pm 1/\sqrt{2}$. We see that, even for the $N=24$ site cluster at hand, we are able to numerically resolve this structure, confirming that the KSL is indeed a non-Abelian quantum spin liquid described by an Ising TQFT. 

\noindent(iii) $h\gg K$. For sufficiently large magnetic field the system will clearly become polarized along the axis of the external field. In this polarized phase (PL) the ground state is a trivial product state and the lowest energy excitations are conventional magnon modes. 

\begin{figure}[!t]  
\includegraphics[width=\columnwidth]{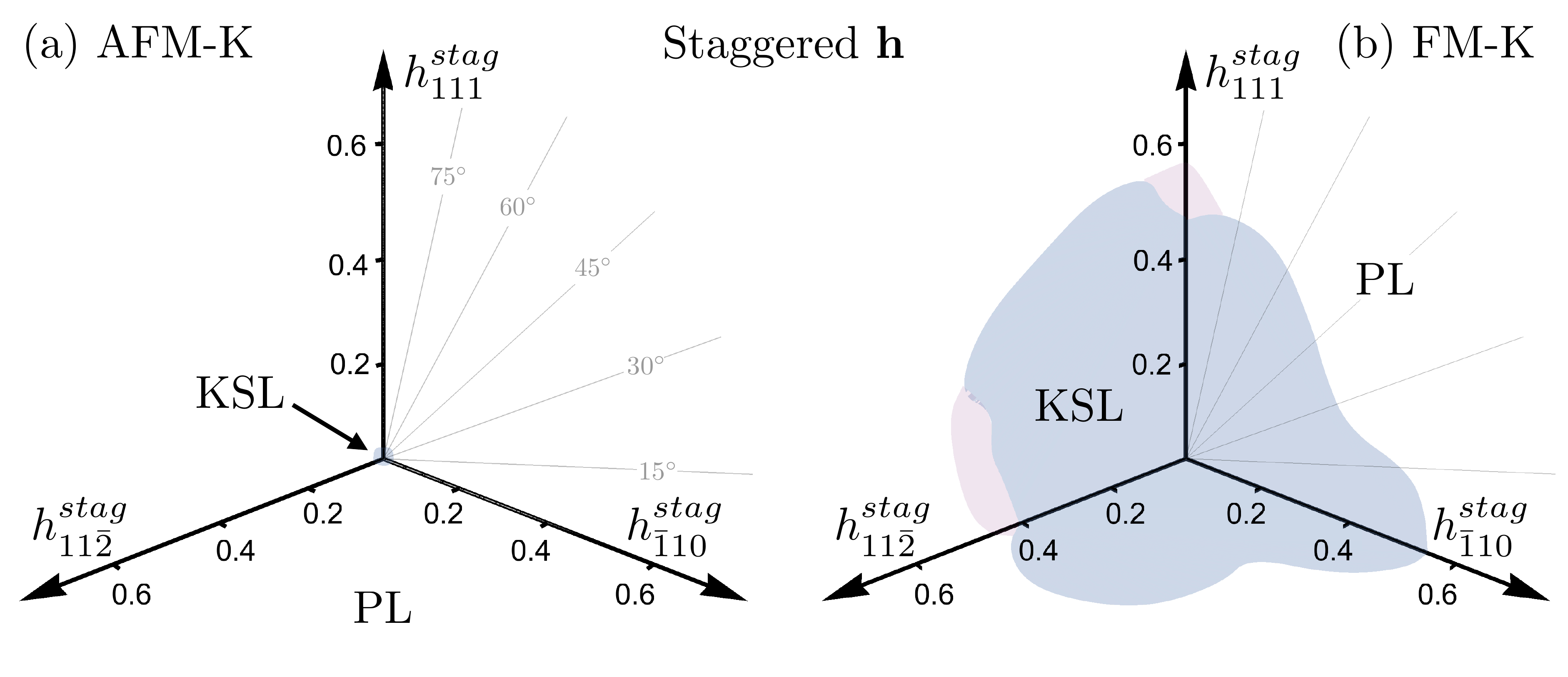}
\caption{Phase diagrams in a staggered magnetic field. (a) The pure AFM Kitaev model and (b) the pure FM Kitaev model.
	     The pink shading marks a region of potential interest (not explored here).}
\label{fig:StagPDs}
\end{figure}

The phase diagrams of Fig.~\ref{fig:UPDs} expand this perspective by providing the critical field strengths, at which the KSL
is destroyed, for tilted field setups.  As can be seen in Fig.~\ref{fig:UPDs} the critical field does not depend sensitively on the field direction (though in real materials anisotropic $g$-factors need to be considered that will distort the phase diagram). What is strikingly evident, however, is that there is a marked contrast in the stability of the KSL in the case of AFM versus FM coupling, with an order of magnitude difference in the critical fields. 
To investigate the source of this difference we show in Fig.~\ref{fig:StagPDs} the phase diagrams for a {\em staggered} external field, with $+\vec{h}$ applied on one sublattice and $-\vec{h}$ on the other sublattice of the honeycomb lattice. We see that, in this case, there is still an order of magnitude difference in the critical fields but now the situation has been reversed. The AFM KSL is significantly less stable in a staggered field compared to a uniform one, while the FM KSL is less stable in a uniform field and significantly more stable in a staggered one. The stability of the KSL thus crucially depends on whether the applied field matches the underlying spin correlations or not. 
We expect this observation to generically hold and to also apply to the three-dimensional generalizations of the Kitaev model \cite{Obrien2016classification} under an external field. Though it is experimentally not possible to generate a staggered field using conventional magnets, it may be possible to realize the desired effect by placing thin samples of a Kitaev material on a substrate which is a trivial honeycomb antiferromagnet, producing a staggered field by proximity, and thereby allowing to probe this effect.  \\

\noindent {\bf Intermediate gapless phase --} Beyond the KSL there is, for a wide range of field angles in the case of AFM Kitaev couplings, an intermediate phase before entering the high-field PL state. To investigate the properties of this phase we focus on two generic cuts away from any high-symmetry directions, shown by the dashed (red) lines in Fig.~\ref{fig:UPDs}, one close to the in-plane $[\bar{1}10]$ direction  at $\theta=7.5^\circ$ ($\pi/24$) and the other close to the out-of-plane $[111]$ direction at $\theta=82.5^\circ$ ($11\pi/24$) (for an example of a cut in which there is a single, direct KSL-PL transition see Supplementary Note $2$).  

\begin{figure}[!t]  
\includegraphics[width=\columnwidth]{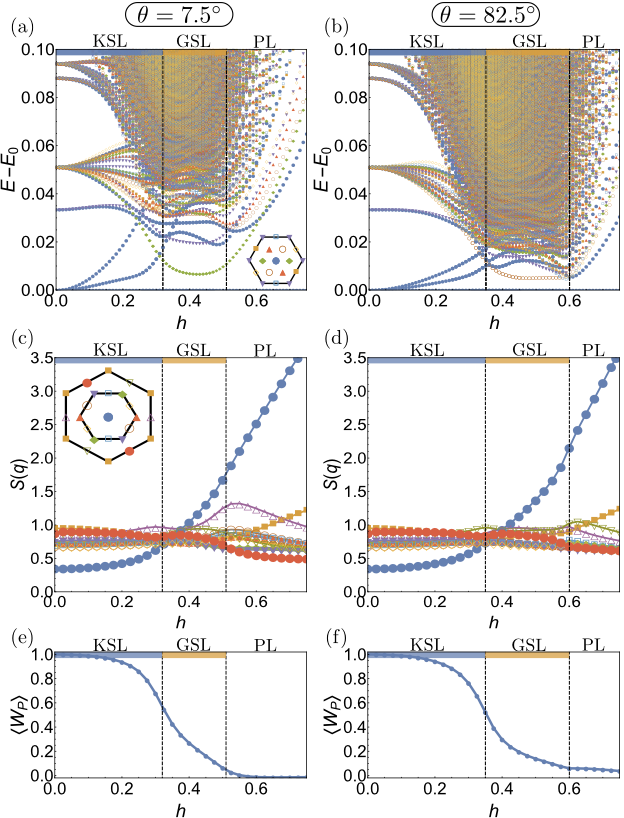}
\caption{Energy spectrum, static spin structure factor, and plaquette flux. In (a) - (c) a cut through the pure AFM Kitaev phase diagram at an angle of $\theta=7.5^\circ$  and in (d) - (f) at $\theta=82.5^\circ$. These cuts correspond to the red dashed lines in Fig.~\ref{fig:UPDs}(a).}
\label{fig:EnergySSF}
\end{figure}

One striking signature for the transition from the KSL to this intermediate phase is a dramatic increase in the density of states at low energies. This is illustrated in Figs.~\ref{fig:EnergySSF}(a), (b), which show the full low-energy spectrum as a function of increasing field magnitude (obtained from numerical exact diagonalization) for the two cuts, with states labelled by their momentum quantum number. Indeed, for the energy window shown here there are more than $10$ times as many states within the intermediate phase as there are in the zero-field Kitaev limit. This increase of the low-energy density of states by more than an order of magnitude (in combination with no degenerate ground-state manifold) is a strong indication that the intermediate phase is in fact a {\em gapless} phase with considerably more low-energy modes than the gapless KSL in the vanishing field limit. This incredible density of states at low energies is supported by finite-size scaling, shown in the Methods section, making it a robust feature of the intermediate phase. Beyond the intermediate phase, we see that this plethora of low-lying states quickly get pushed up linearly, consistent with the notion that a spin gap begins to open upon the transition to the PL state.

To probe the magnetic nature of this intermediate gapless phase, we turn to the static spin structure factor of the ground state, 
$S(\vec{q})=\frac{1}{N} \sum_{i,j} \avg{\vec{S}_i \cdot \vec{S}_j} e^{i\vec{q}\cdot\left(\vec{r}_i - \vec{r}_j\right)}$,
which is plotted in Figs.~\ref{fig:EnergySSF}(c), (d) for the two cuts.
There are no clear signs of any magnetic ordering, with only the $\Gamma$ point intensity (i.e.~the magnetization) significantly changing as the transition from the KSL to the intermediate phase is crossed. The flat, rather featureless structure factor of the intermediate phase is indicative of a quantum spin liquid phase. We also show the flux of the $\intg_2$ gauge field through the plaquettes of the honeycomb lattice, $\avg{W}_P = \avg{S_i^x S_j^y S_k^z S_l^x S_m^y S_n^z}$, in Figs.~\ref{fig:EnergySSF}(e), (f). This quantity does not show visible signatures of the transitions. The flux $\avg{W}_P \approx 1$ in the KSL phase and $\avg{W}_P \approx 0$ in the PL phase. In the intermediate phase it takes a range of intermediate values, interpolating between these two limits. This indicates that the plaquette flux is heavily fluctuating in the intermediate phase. Taken together, all of these results are consistent with a gapless, disordered state, allowing us to identify the intermediate phase as a gapless spin liquid (GSL).  

\begin{figure}[!t]  
\includegraphics[width=\columnwidth]{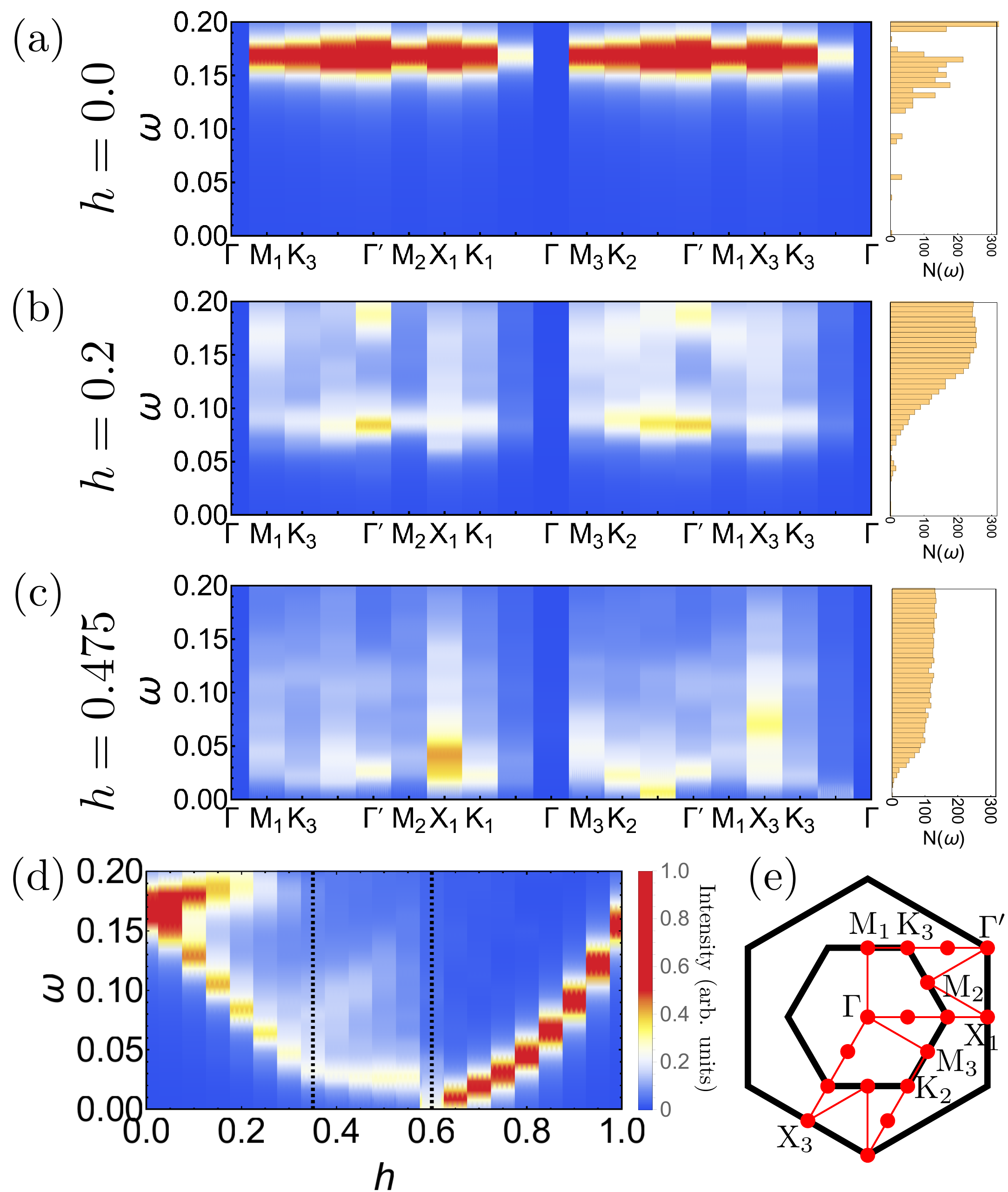}
\caption{Dynamical spin structure factor. (a) The zero-field KSL, (b) a point midway in the KSL phase 
		and (c) a point in the middle of the GSL 
		along a path through all high-symmetry points of the extended Brillouin zone as illustrated in (e). In (d) the intensity at the $\Gamma^\prime$ point is shown as a function of increasing field 
		along the cut $\theta=82.5^\circ$, the upper of the two dashed red lines in Fig.~\ref{fig:UPDs}(a).}
\label{fig:DySF}
\end{figure}

\begin{figure}[!t]  
\includegraphics[width=\columnwidth]{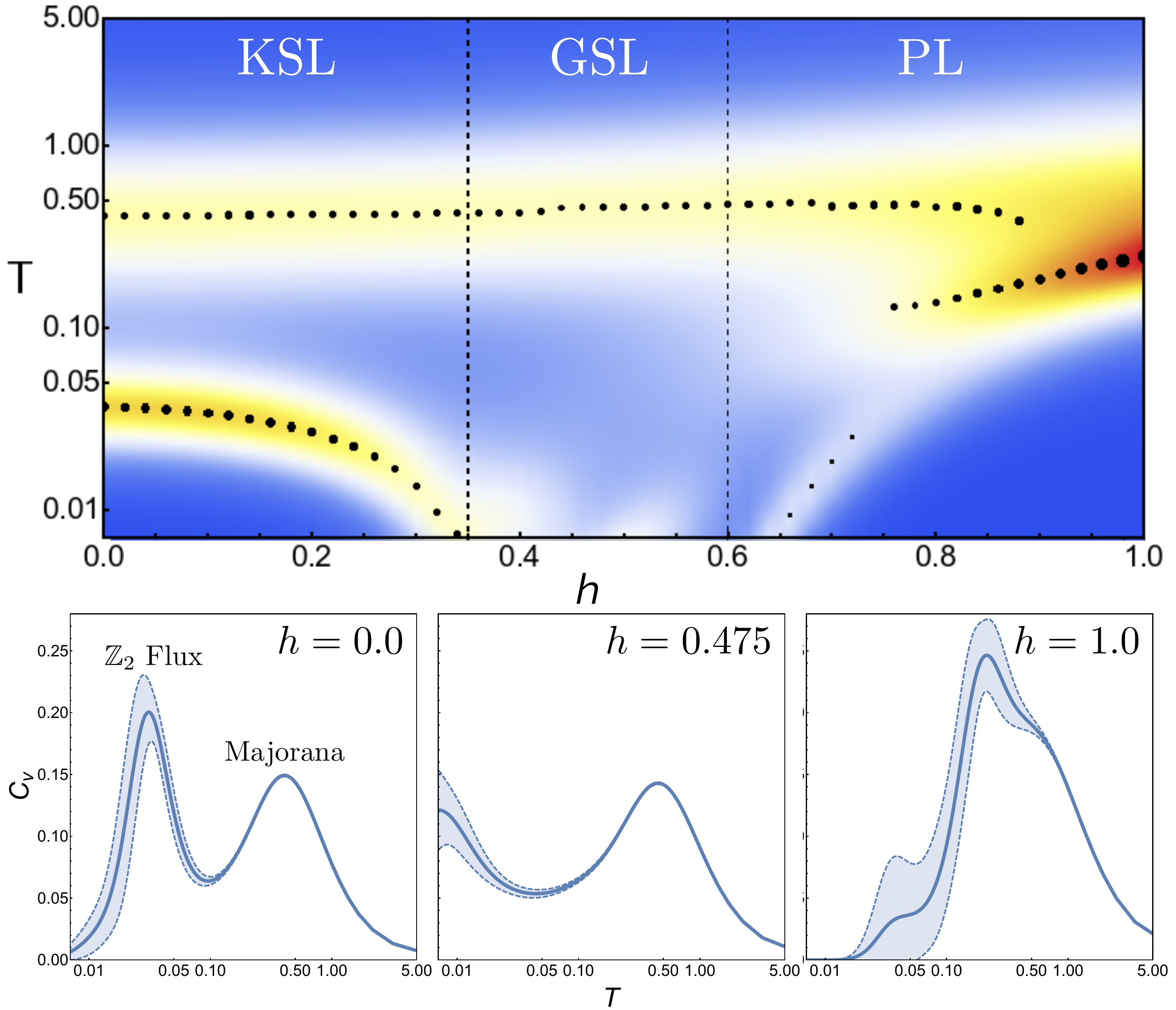}
\caption{Specific heat as a function of temperature for increasing field. Shown along the cut at $\theta=82.5^\circ$ (upper panel).
		The black circles (and their widths) indicate the location (and heights) of the maxima.
		The three lower panels show specific heat scans for the zero-field KSL, the intermediate GSL at $h=0.475$ (note that though it is not shown here, the specific heat goes to zero as $T\rightarrow 0$),
		and the PL state at $h=1.0$, respectively. The light blue shading indicates the standard deviation of the estimates.}
\label{fig:Cv}
\end{figure}

\begin{figure*}[!t]  
\includegraphics[width=\textwidth]{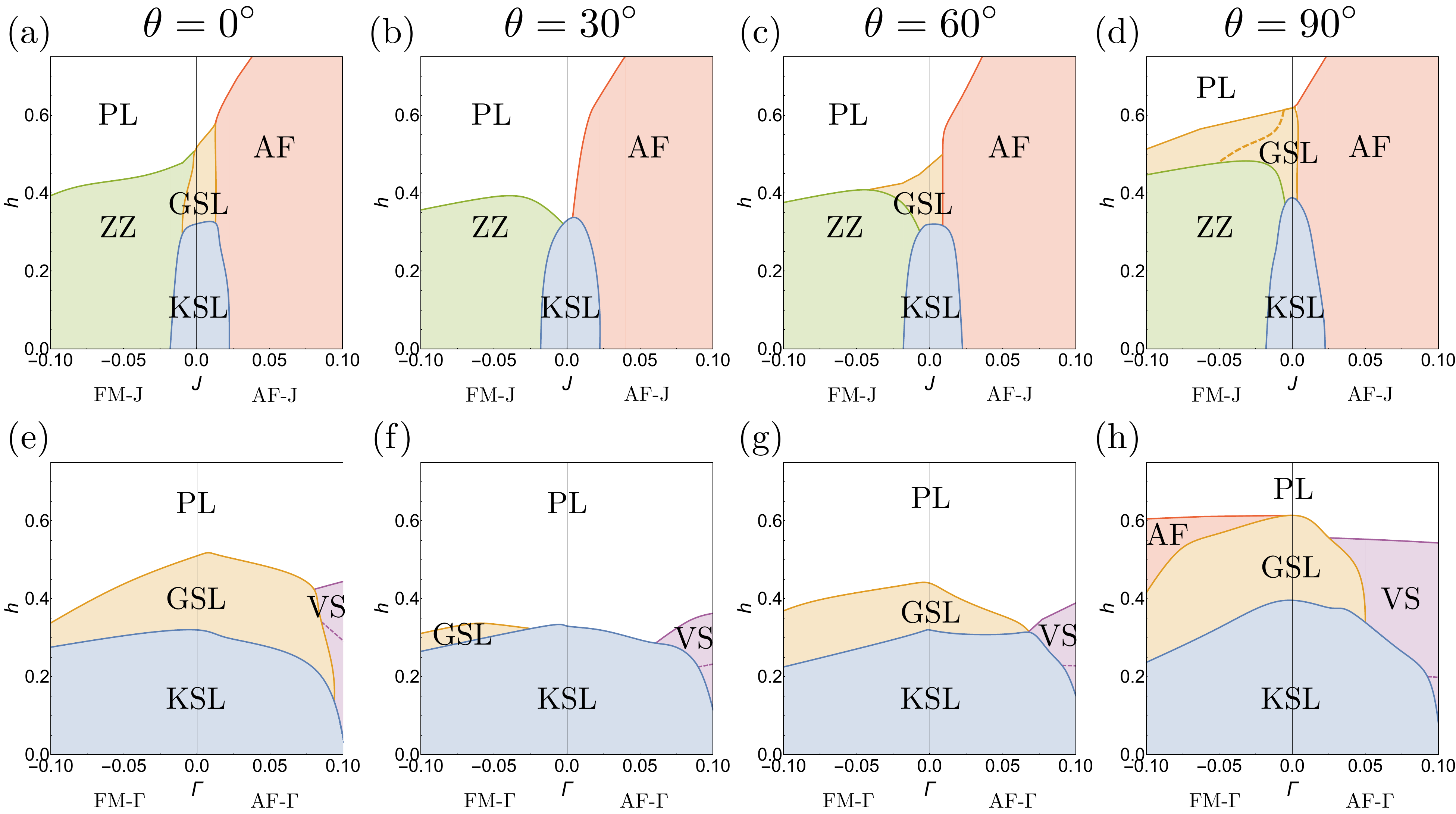}
\caption{Phase diagrams beyond the pure Kitaev model. Effects of (a-d) an additional Heisenberg and 
		(e-h) an additional off-diagonal $\Gamma$ exchange on the KSL and GSL in tilted magnetic fields. The Kitaev interaction is parameterized by $\alpha$ and the perturbing coupling by $\left(1-\alpha\right)/2$, with $\alpha=1$ corresponding to the pure Kitaev model and $\alpha=0.8$ to the edges of the phase diagram, $J=\pm0.1$, $\Gamma=\pm0.1$. In the presence of additional Heisenberg interactions both spin liquid phases eventually give way to
		either zig-zag (ZZ) or antiferromagnetic (AF) ordering. 
		The inclusion of the additional off-diagonal $\Gamma$-exchange affects the stability of the spin liquid phases
		in a lesser way, with AF or vortex state (VS) order arising only for large coupling strengths.
		Note that in all phase diagrams the GSL appears to piggypack on the KSL phase, indicating that it is in fact
		of a descendant of the KSL. 
		 }
\label{fig:PertPDs}
\end{figure*}

This immediately raises the question about the origin and nature of the gapless degrees of freedom. To answer this question it has proved particularly insightful to look at the the dynamical spin structure factor, which provides strong indications that it is the vison gap which closes at the transition to the intermediate gapless phase. The dynamical spin structure can be written in Lehmann representation as
\begin{equation}
S^{\alpha\alpha} \left( \vec{Q}, \omega\right) = \sum_{n} | \bra{n} S^\alpha_\vec{Q} \ket{0} |^2 \delta\left( \omega - (E_n - E_0)  \right) ,
\end{equation}
where we note that the $n=0$ contribution, which, for $\vec{Q}=\Gamma$, is simply $|\avg{S^\alpha_{\rm total}}|^2$, the magnetization induced by the external field, is not included in the following discussion. Furthermore, from now on, we will focus on the sum $S \left( \vec{Q}, \omega\right)  = \sum_\alpha S^{\alpha\alpha}\left( \vec{Q}, \omega\right)$. At zero field there is, despite the system being gapless in this limit, a distinct gap to physical spin excitations as these involve the creation of gapped $\intg_2$ flux excitations \cite{Knolle2014dynamics,Knolle2015dynamics}. This flux gap is clearly visible in Fig.~\ref{fig:DySF}(a), with its uniformity across momenta reflecting the static nature of the flux excitations. Note that the flux gap is absent at the $\Gamma$ point for the AFM Kitaev model due to the AFM correlations of the ground state. Upon applying the magnetic field, this uniform flux gap breaks apart and a significant portion of the spin spectral weight is pushed to zero energy across the whole Brillouin zone as illustrated in Fig.~\ref{fig:DySF}(b)
for a point midway in the KSL phase along the cut at $\theta=82.5^\circ$. These states are further pushed down in energy as the transition to the intermediate phase is crossed, with Fig.~\ref{fig:DySF}(c) showing results for a point in the middle of the intermediate phase.
The overall spin spectral weight of these low-energy states makes up a significant part ($\sim 40\%$) of the zero-field flux gap. This is strong evidence that the transition from the KSL to the intermediate phase is thus marked by the closure of the flux gap (see Supplementary Note $3$ for further discussion of the flux gap closure for the cut at $\theta=7.5^\circ$). In the intermediate phase the dynamical structure factor at higher energies remains featureless, with weight distributed across all energies. There are no signatures of pseudo-Goldstone modes or any kind of conventional magnon excitations. These features support the case for a gapless quantum spin liquid
arising from the closing of the $\intg_2$ flux gap.

The key role played by the flux excitations is also visible at finite temperatures. In Fig.~\ref{fig:Cv} we show the specific heat as a function of increasing field, calculated using the method of thermal pure quantum states \cite{TPQS2012,CanTPQS2013}. At zero field, it has been established through numerical exact Monte Carlo simulations \cite{Nasu2014vaporization} that there are two finite temperature crossovers, a high-temperature one associated with the itinerant Majorana fermions indicating the fractionalization of the original spins and a low-temperature one associated with the $\intg_2$ gauge field, at which it orders into its ground state configuration. The location of these peaks is
correlated to the bandwidth of the fermion hopping and the vison gap in the gauge sector respectively. As one approaches the transition to the intermediate phase the low-$T$ peak starts to drift to lower and lower temperatures. This is another telling sign that the energy scale associated with the gauge field is lowered as the field increases. Interestingly, the high-$T$ peak does not show any notable changes as the transition to the intermediate phase is crossed. This would seem to suggest that the itinerant Majoranas are not affected by the transition, with all of the action occurring only in the gauge sector. Once the PL phase is entered a single peak develops, as expected since fractionalization is lost. \\


\noindent {\bf Beyond the Kitaev model --} Before turning to a discussion of the nature of the intermediate gapless spin liquid phase, we round off our numerical results with a study of its stability in the presence of microscopic perturbations. In any Kitaev material the bond-directional exchanges 
of the Kitaev model are accompanied by other, more conventional, interactions such as symmetric off-diagonal exchange terms 
($\Gamma$-interactions) along with an isotropic Heisenberg coupling \cite{Rau2014,Rau2016,Trebst2017}. In Fig.~\ref{fig:PertPDs} we illustrate phase diagrams elucidating the effects of these additional couplings in the vicinity of the pure Kitaev model for various
tilt angles of the magnetic field. It is clear that the intermediate GSL is stable under both kinds of perturbations. Indeed the stability of the GSL mirrors that of the KSL, always sitting above it in these phase diagrams. This suggests that the intermediate phase is in fact a descendent of the KSL, resulting from an instability of the gapped Ising anyon phase. \\


\noindent{\bf Discussion --} To summarize our key results, we have established that the Kitaev honeycomb model contains another phase exhibiting unconventional magnetism alongside its already well-known gapless and gapped $\intg_2$ spin liquid phases. This additional phase is gapless with a dense continuum of excitations, featureless structure factor (both static and dynamic), 
fluctuating $\intg_2$ fluxes, and low-energy spin spectral weight. 

To reveal the precise nature of this phase the central question is what its gapless degrees of freedom are -- matter or gauge fields or both? Though this question is difficult to answer definitively, we interpret our results as providing multi-faceted evidence for the emergence of a $U(1)$ spin liquid, in which gapless fermions are coupled to a massless gauge field. 
Our conclusion is guided by the following observations.

(i) The energy spectrum shows that the transition from the KSL to the GSL occurs through the dramatic shift of a large density of states to low energies, forming a dense continuum of gapless excitations. The dynamical structure factor reveals that these states carry with them physical spin spectral weight at all momenta. Since we know from exact studies at zero field that only the vison excitations of the KSL (but not the Majorana fermions) carry spin spectral weight, this indicates,
supported also by the thermodynamic signatures in the specific heat, 
that this transition is marked by the closing of the vison gap, resulting in a massively fluctuating gauge field. Naively, one could identify this gap closing with vison condensation which in turn should lead to confinement and a trivial, magnetically ordered phase (such as the transition from the KSL to zig-zag magnetic ordering in the presence of additional Heisenberg interactions \cite{Chaloupka2013}). However, in the KSL, the visons carry a Majorana zero mode (which is the hallmark of its topological order) and so cannot condense by themselves. If the visons indeed avoid condensation at the gap closing transition, this leads to an intriguing scenario in which their associated Majorana fermions form a gapless Fermi surface coupled to a massless gauge field.

(ii) A particularly enlightening and intuitive perspective on the transition to the GSL can be gained by considering an alternative Abrikosov fermionic parton decomposition of the spin operators. Such partons naturally possess an accompanying $U(1)$ gauge structure. If we imagine starting from a phase in which the fermionic partons, coupled to such a $U(1)$ gauge field, form a Fermi surface the KSL can naturally be accessed through a pairing instability of the fermions \cite{Metlitski2015,Galitski2007}. The formation of a superconducting condensate Higgses the gapless $U(1)$ gauge field down to a gapped $\intg_2$ gauge field. In order to properly match the topological properties of the KSL, the superconductor must be a chiral $p$-wave superconductor, which ensures, for example, that flux excitations can bind Majorana fermions \cite{NayakParton2011}. Starting from zero field, we see that, in this picture, the transition from the KSL to the intermediate GSL can be understood as a transition from a gapped chiral $p$-wave superconductor, coupled to a gapped $\intg_2$ gauge field, to a gapless spinon Fermi surface, coupled to an emergent $U(1)$ gauge field. The Fermi surface can be stabilized by a lack of lattice symmetries and momentum conservation \cite{MotrunichFS2011}.

The key to the realization of this scenario in the present context is that we can have an {\em emergent} $U(1)$ conservation of the fermionic partons, with the closure of the vison gap thus related to the emergence of this conservation law. The Majoranas remain intact throughout the transition, with the only change being that, in the GSL, the Majoranas can now be combined into complex fermions with an emergent $U(1)$ particle conservation. This naturally explains why the Majorana peak in the specific heat is relatively unaffected as the transition to the GSL is crossed. 

\begin{figure}[]  
\includegraphics[width=\columnwidth]{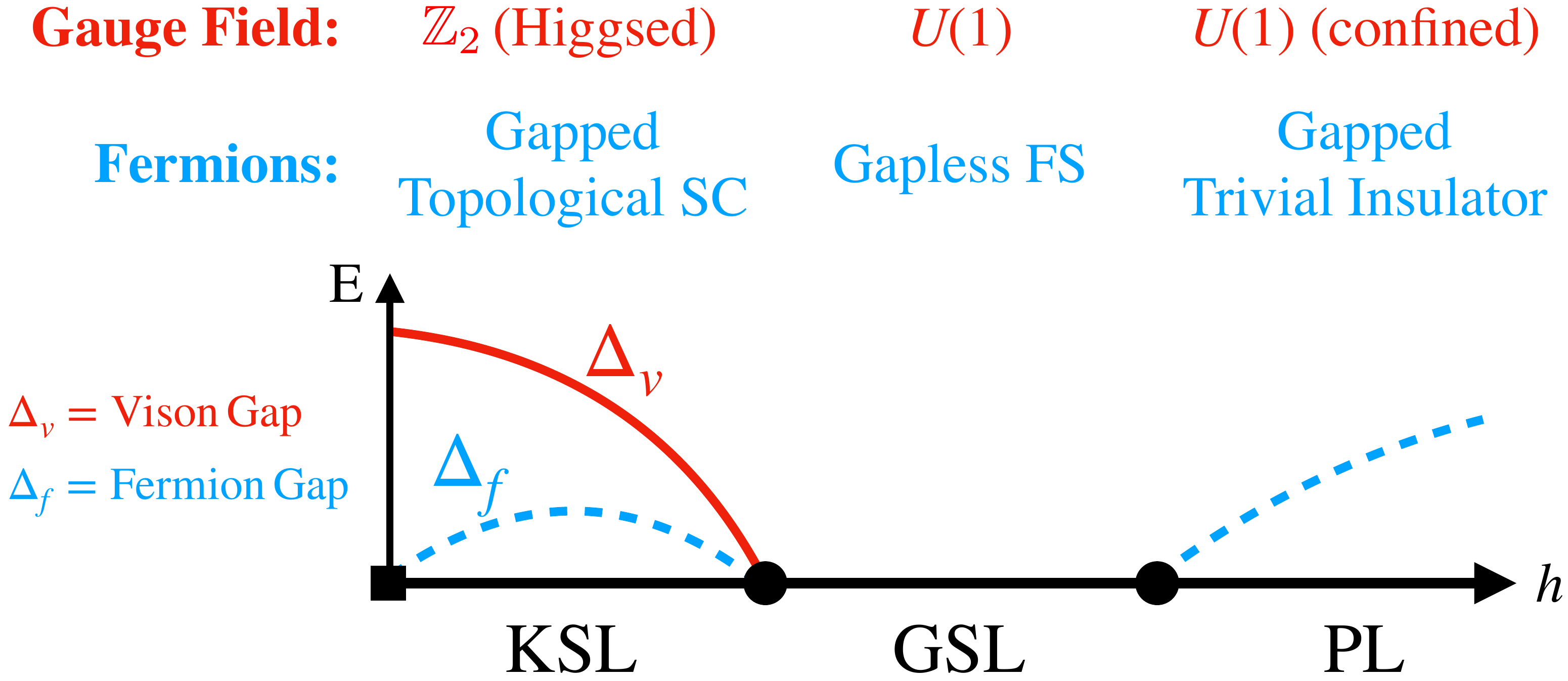}
\caption{Schematic phase diagram from the perspective of fermionic partons. The behavior of the fermions and associated gauge field indicated for the KSL, GSL and PL phases. The flux (vison) gap $\Delta_v$ and the fermion gap $\Delta_f$ is also shown.}
\label{fig:FT}
\end{figure}

The transition from the GSL to the PL phase can be similarly understood within this framework, corresponding to a transition from a gapless Fermi surface to a gapped trivial insulator. With the fermions completely gapped, they can be integrated out to produce a low-energy theory of a pure compact U(1) gauge theory. However, such a theory is well-known to be unstable to confinement via monopole proliferation \cite{PolyakovCQED}, resulting in a completely trivial gapped phase, the PL phase. The complex Abrikosov fermion perspective thus naturally gives an intuitive and unified description of all of the numerical data at hand. This is summarised in Fig.\ref{fig:FT} in which the behavior of the gauge field and fermionic partons in the KSL, GSL and PL phases is detailed. 

(iii) This parton perspective also motivates a natural field theory description of the transition from the KSL to the GSL in terms of a Fermi surface of partons coupled to a dynamical $U(1)$ gauge field 
\begin{equation}
\mathcal{L} \!= \! \psi^\dagger \!\left( D_t \!-\!\frac{1}{2m} D_i^2 \!-\! \mu \right) \!\psi + g \left( \psi^\dagger \!\Delta \psi^\dagger + h.c. \right)+  \dots , \label{eqn:FT}
\end{equation}
where $\psi$ represent the low-energy fermionic parton modes near the Fermi surface, $D_\mu = i \partial_\mu +  a_\mu$, $\Delta = \abs{\Delta} \left( \partial_x \pm i \partial_y \right)$ corresponds to a decoupling in the chiral p-wave channel of an attractive four-fermion interaction of strength $g$ and the higher order terms include a kinetic term for the dynamical gauge field $a$. For $\Delta=0$ we have a Fermi surface of partons coupled to a dynamical U(1) gauge field. This corresponds to the GSL phase. Though normally one might expect such a Fermi surface to be immediately susceptible to a pairing instability, here we have an extended phase, with the Fermi surface stable up to a finite critical strength of the atttractive interaction \cite{Metlitski2015} and further stabilized by the lack of lattice symmetries and momentum conservation discussed above (see Supplementary Note $4$ for further discussion). At the transition pairing onsets, such that for $\Delta\neq 0$ the fermions become gapped and can be integrated out. If the gapped fermions occupy a topologically non-trivial band structure this will generate a Chern-Simons term at level $1/2$, ensuring that in vortex cores (where the superconducting condensate vanishes) there is a bound Majorana fermion \cite{NayakParton2011}, and a mass term for the gauge field generated by the Anderson-Higgs mechanism. This corresponds to the KSL phase. This simple, minimal theory is thus able to capture the physics either side of the transition. 

(iv) Lastly, let us mention, for completeness, an alternative scenario for the transition between the KSL and GSL. The scenario, which we can definitively rule out based on our numerical data, starts from the KSL and argues \cite{Lahtinen2012,Lahtinen2014} that instabilities of this topological phase can be driven by the condensation of Ising anyons, which are brought into close proximity with increasing field strength. However, the ensuing phase is still a chiral spin liquid (albeit with an Abelian topological order) which would reveal itself through a ground-state degeneracy that we do not observe in our numerical data for the GSL (in contrast to the KSL where the three-fold quasi-degenerate ground states strongly corroborate its topological nature via the $S$-matrix calculation showcased above).

In conclusion, our numerical analysis of the complete phase diagram of the Kitaev model in tilted magnetic fields has revealed that this fundamental model harbors not only $\intg_2$ spin liquid physics, but also exhibits an extended spin liquid regime with a distinct $U(1)$ gauge structure. Our numerical observation of the phase transition between these two regimes at finite field strengths provides a multi-faceted perspective of the accompanying signatures in static, dynamical, and finite temperature quantities. 
It will be an interesting avenue for future theoretical studies to further investigate the field theory description for this transition, which clearly lies beyond the standard
Landau-Ginzburg-Wilson paradigm. Though currrent Kitaev materials are all believed to possess FM Kitaev couplings, the possibility of an AFM coupling is not ruled out on any miscroscopic grounds (early reports suggested that RuCl$_3$ possessed exactly such an AFM coupling) and indeed there is recent work suggesting that they may naturally appear in $f$-electron based systems \cite{fMotome2018}. Future experimental studies on such Kitaev materials might be able to probe the nature of the fractional excitations in the gapless spin liquid regime and reveal the existence of a Fermi surface. \\


\noindent {\bf Acknowledgements --} 
We thank A. Rosch for useful discussions. This work was supported by the Deutsche Forschungsgemeinschaft (DFG, German Research Foundation) - Projektnummer 277101999 - TRR 183 (project B01). The numerical simulations were performed on the CHEOPS cluster at RRZK Cologne and the JUWELS cluster at FZ J{\"u}lich. \\


\noindent{\textbf{Methods:}} 
{\small
\noindent {\it Exact diagonalization and finite-size scaling --} The exact diagonalization results were produced using the library 
ARPACK \cite{arpackusers}, primarily on an $N=24$ site cluster with the full point group symmetry of the honeycomb lattice, and 
containing all the high symmetry points of the Brillouin zone. Additional calculations were done on system sizes ranging from 
$N=18$ to $N=32$ sites, with qualitatively consistent results. 

For a point mid-way in the GSL phase we show in Fig.~\ref{fig:FSS}(a) the energy difference between the ground state 
and the lowest lying state from each momentum sector for $N=18, 20, 24, 28, 30$ and $32$ site clusters. The $N=18, 24$ and $32$ site 
clusters are highlighted in red as these are are the only clusters that have the full point group symmetry of the honeycomb lattice. 
Unfortunately, unlike the N=24 site system, the $N=18$ and $32$ site clusters do not contain all of the high-symmetry points in the 
BZ, marking out the $N=24$ site cluster as unique and why we chose to show data for this cluster in the main manuscript. For the 
largest system sizes the density of states at low-energies increases with increasing system size, with the largest gap for any momentum 
sector in the $N=32$ site case being just $0.004 K$. We also use a solid (red) line to indicate the gap to the first excited state for the 
symmetric clusters. The finite-size scaling is clear evidence that the intermediate phase is gapless, with an incredibly dense spectrum of 
excited states at low energies from all momentum sectors (for a detailed comparison to other ED studies see Supplementary Note $5$).  

\begin{figure}[!b]  
\includegraphics[width=\columnwidth]{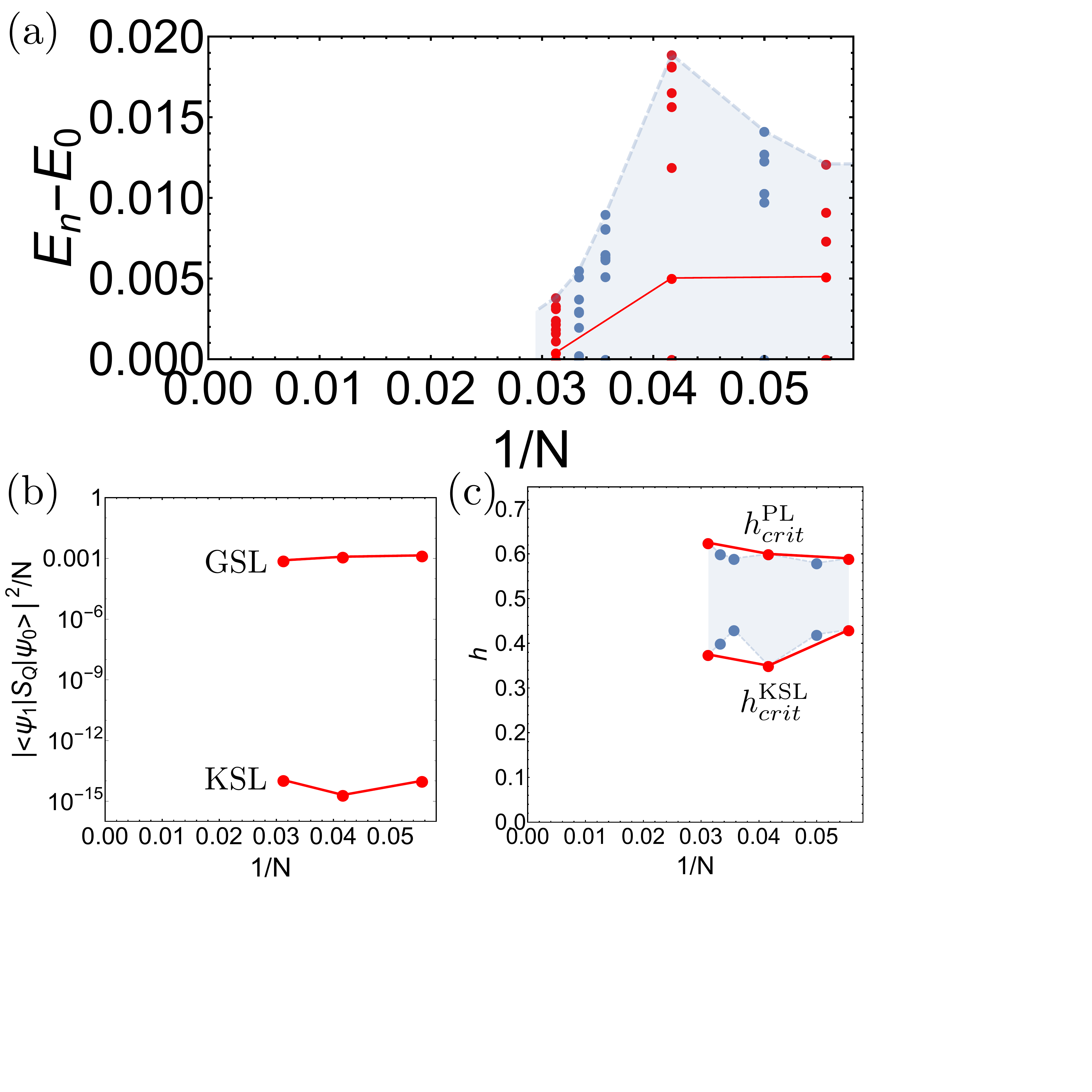}
\caption{Finite-size scaling. Shown, for a range of system sizes, are 
(a) the energy gaps between the ground state and the lowest lying state from each momentum sector for a 
point mid-way in the GSL phase, 
	(b) the critical fields associated with the transition out of the KSL and into the 
PL phase and 
	(c) the spin spectral weight associated with the first excited state in the KSL and GSL phases. All data is taken along the cut corresponding to the upper of the two dashed lines (red) in Fig.~\ref{fig:UPDs}(a).}
\label{fig:FSS}
\end{figure}

We also show the spin spectral weight associated with the first excited state, $\sum_\alpha | \bra{1} S^\alpha_\vec{Q} \ket{0} |^2 $, in both the KSL and GSL phases for the three symmetric 
clusters in Fig.~\ref{fig:FSS}(b) (for the KSL we choose the first excited state above the ground state degenerate manifold). In the KSL phase, since all of the spectral weight is concentrated above the finite flux gap 
the weight associated to the first excited state is practically zero. On the other hand, at the same point mid-way in the GSL phase, the flux gap has collapsed to zero 
with the first excited state now showing finite spin spectral weight. This clearly demonstrates that the field-induced closure of the flux gap, via 
the transfer of spin spectral weight to zero energy, is a robust feature of the intermediate phase.

Finally we show the critical fields associated with the transition out of the KSL, $h_{crit}^{KSL}$, and the transition into the 
PL phase, $h_{crit}^{PL}$, in Fig.~\ref{fig:FSS}(c). This clearly indicates that the intermediate phase is stable and its size is roughly $\sim0.2 K$ for the cut shown.  \\

\noindent {\it Determination of the phase boundaries --} The phase boundaries for the phase diagrams presented in the main text were determined using a combination of the second derivative of the ground state energy and the ground state fidelity, taken from a range of radial cuts (26 in
angular spacings of $\pi$/100 and radial field spacing of 0.01/0.001 for the AFM/FM cases, a
total of 1976 points) for each of the three 2d phase diagrams presented in Figs.~\ref{fig:UPDs} and \ref{fig:StagPDs}
(i.e. a total of 6000 parameter points were computed for each of Fig.~\ref{fig:UPDs}(a), (b), Fig.~\ref{fig:StagPDs}(a), (b)). For the ground state energy, it is a peak in its second derivative which indicates the presence of a phase transition. The ground state fidelity is defined as $F(g) = \braket{\Psi_0(g) | \Psi_0(g+\delta g)}$ for some tuning paramter $g$ (in our case the magnitude of the magnetic field $h$). A first order transition, i.e.~a level crossing, is signified by a discontinuity in the fidelity, while a second order transition results in a smooth dip. For the two cuts focused on in the main text, $\theta=7.5^\circ$ and $\theta=82.5^\circ$ in the $c$-$b$ plane, we show in Fig.~\ref{fig:SMPB} these two quantities as a function of field magnitude. Two transitions can clearly be resolved, with excellent agreement between the two distinct quantities. Finally, we note that we also find excellent agreement between the phase boundaries computed for our $N=24$ site cluster and the phase boundaries reported in a recent infinite density matrix renormalization group study of the Kitaev model in a $[111]$ magnetic field \cite{PollmannAFMK2018}. \\

\begin{figure}[!t]  
\includegraphics[width=\columnwidth]{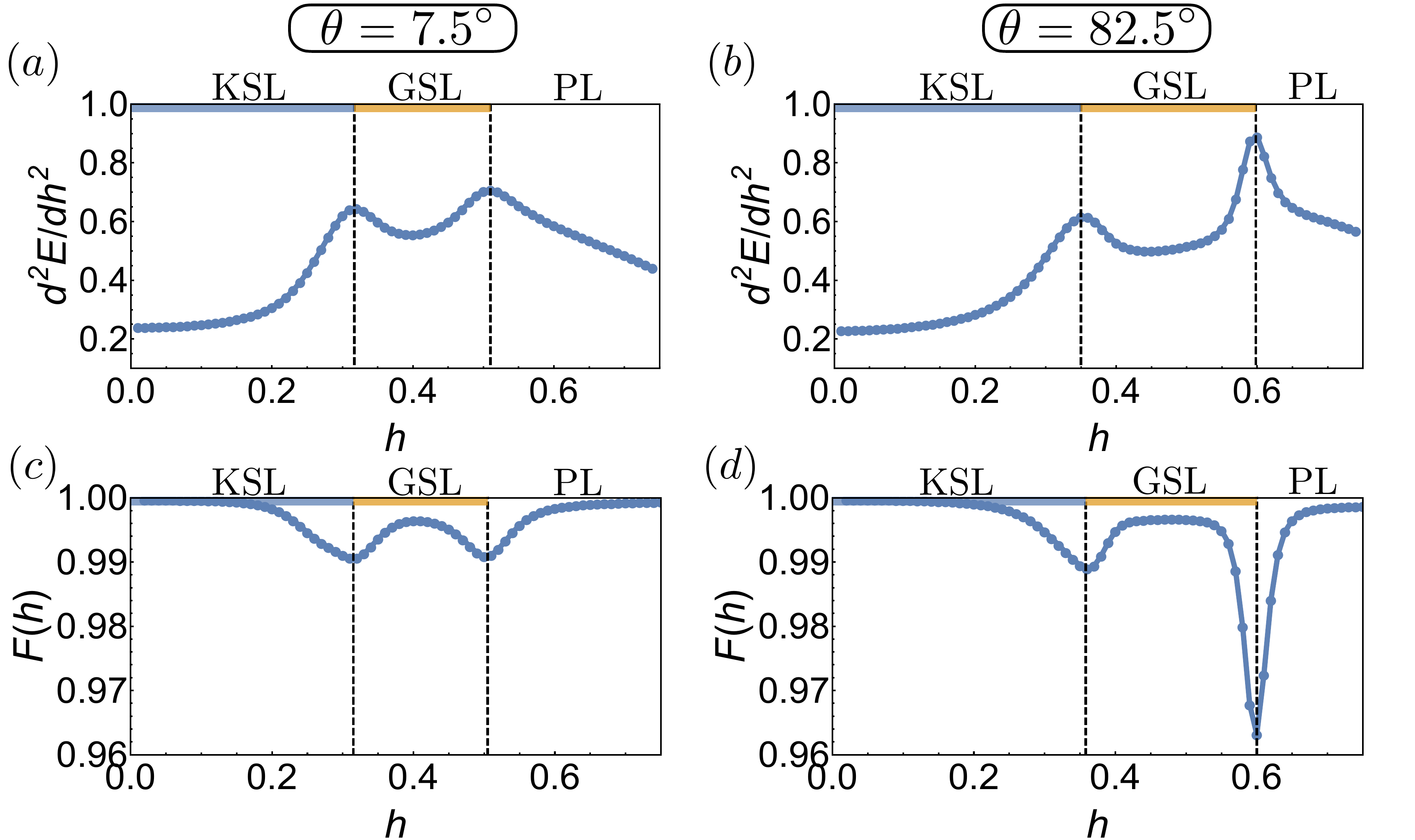}
\caption{Clear numerical signatures of phase transitions. In (a), (b) the second derivative of the ground state energy and (c), (d) the ground state fidelity is shown. These cuts correspond to the dashed lines (red) in Fig.~\ref{fig:UPDs}(a).}
\label{fig:SMPB}
\end{figure}

\noindent {\it Note Added:} During completion of this manuscript a related preprint was posted \cite{Nasu2018}, which studies the phase diagram of the Kitaev model in a $[001]$ magnetic field using a Majorana mean-field approximation. While that work does not discuss the underlying gauge structure of the intermediate phase, it proposes a transition of the Majorana spectrum from gapless Dirac nodes to a nodal line (Majorna Fermi surface) structure -- a scenario reminiscent, but distinct from the (spinon) Fermi surface physics of the $U(1)$ spin liquid put forward in this manuscript.}



%

\pagebreak
\widetext

\setcounter{section}{0}
\setcounter{equation}{0}
\setcounter{figure}{0}
\setcounter{table}{0}
\makeatletter
\renewcommand{\theequation}{S\arabic{equation}}
\renewcommand{\figurename}{}
\renewcommand{\thefigure}{\!S\arabic{figure}}


\section{Supplementary Note 1: Ferromagnetic Kitaev Phase Diagram}

\begin{figure}[!b]  
\includegraphics[scale=0.4]{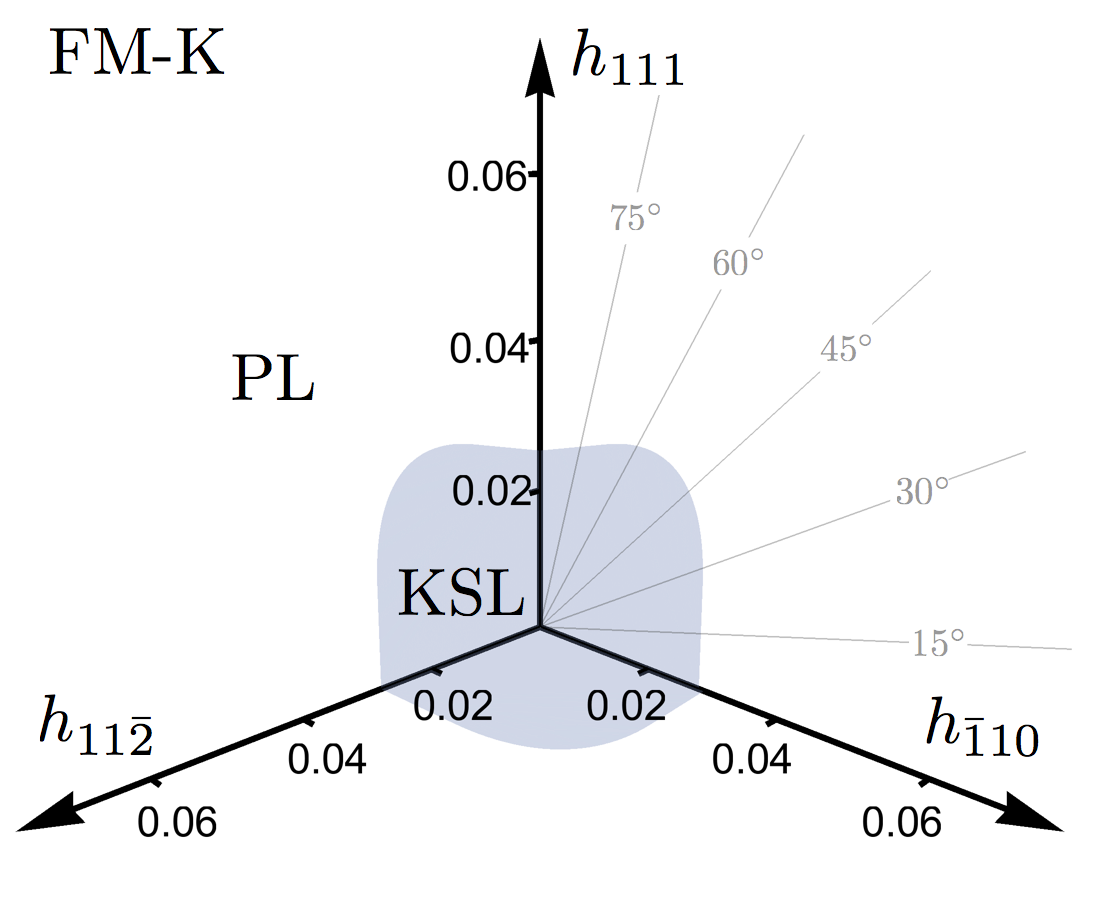}
\caption{Phase diagram in uniform magnetic field. For the pure FM Kitaev model, shown here, there are just two phases, the KSL and PL phases, and a single transition between them.}
\label{fig:SMFM}
\end{figure}

For ferromagnetic Kitaev exchange the phase diagram in tilted magnetic fields exhibits just two phases, the gapped KSL and the trivial PL phase. The phase diagram is shown in the main text in Fig.~\ref{fig:UPDs}(b). In Fig.~\ref{fig:SMFM} we provide the same phase diagram, but with the axes scales reduced by an order of magnitude so as to make the boundaries of the KSL phase more visible. 

\section{Supplementary Note 2: Direct KSL-PL Transition}

\begin{figure}[!t]  
\includegraphics[scale=0.4]{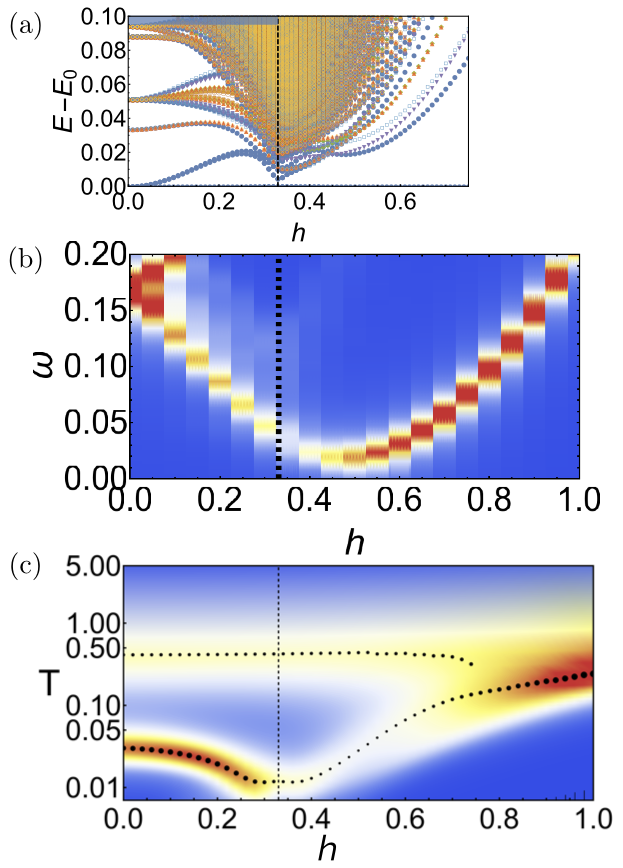}
\caption{Direct KSL to PL transition. Data with (a) the energy spectrum, (b) the dynamical spin structure factor at the $\Gamma^\prime$ point, and (c) the specific heat shown for a cut $30^\circ$ away from $[\bar{1}10]$ toward $[111]$.}
\label{fig:SM30}
\end{figure}

Along certain field directions there is a direct transition from the KSL to the PL phase, with no intermediate GSL phase. In Fig.~\ref{fig:SM30} we show a selection of data for such a scenario. In particular we take a cut at $30^\circ$ away from the $[\bar{1}10]$ direction, within the honeycomb plane, toward the $[111]$ direction, perpendicular to the plane. We show the energy spectrum, the dynamical structure factor for the $\Gamma^\prime$ point and the specific heat as a function of increasing field magnitude. 

\section{Supplementary Note 3: Dynamical Structure Factor}

\begin{figure}[!t]  
\includegraphics[scale=0.4]{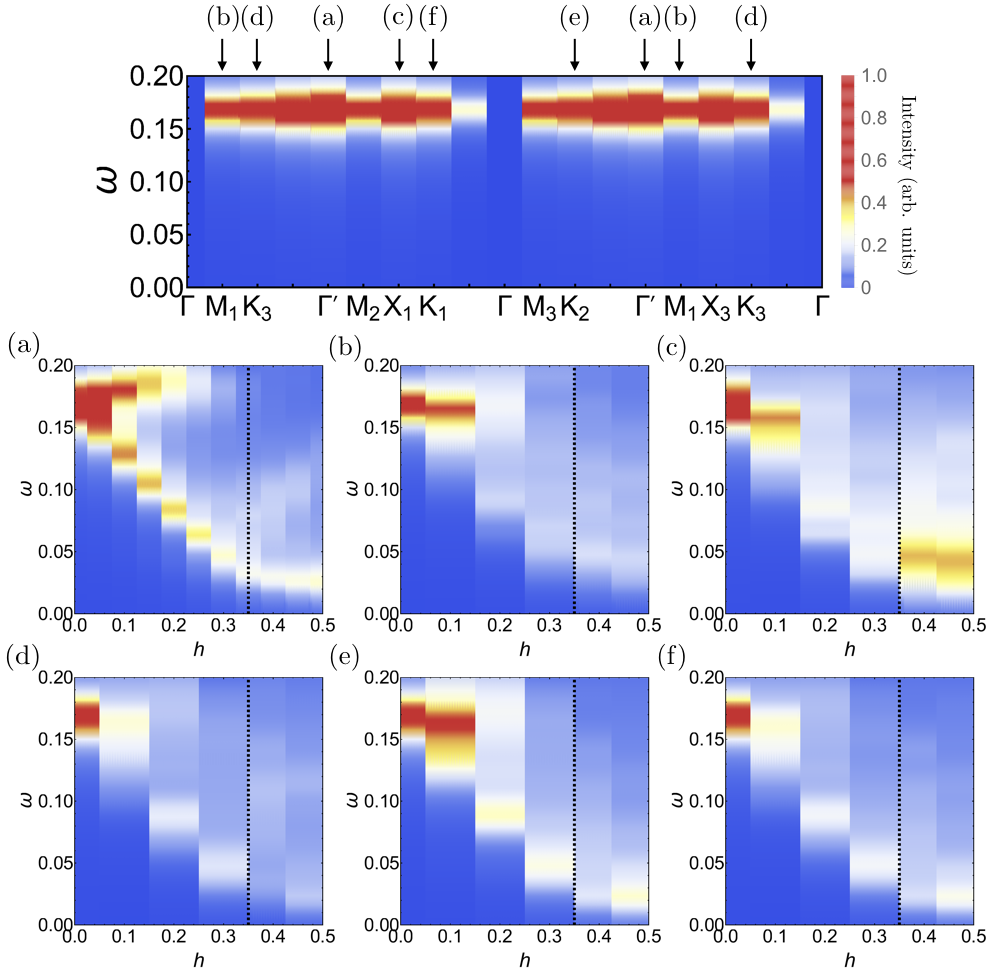}
\caption{Dynamical spin structure factor at various high-symmetry points. (a) $\Gamma^\prime$, (b) one of the $M$ points, (c) one of the $X$ points, and (d)-(f) the three distinct $K$ points. The intensity is shown as a function of increasing field along the cut $\theta=82.5^\circ$, the upper of the two dashed red lines in Fig.~\ref{fig:UPDs}(a) of the main text.}
\label{fig:BzDySF}
\end{figure}

In the main text, it has been argued that the flux gap at zero field breaks apart and moves to lower energy as the external magnetic field is increased, eventually closing fully as the transition to the GSL is crossed. As an example we showcased, in Fig.~\ref{fig:DySF}(d) of the main manuscript, the intensity at the $\Gamma^\prime$ point as a function of field magnitude, with two transitions clearly visible as the spin gap closes at the KSL-GSL transition, remains closed throughout the GSL phase, and then reopens again at the GSL-PL transition. In Fig.~\ref{fig:BzDySF} we show similar plots for other high-symmetry points, namely examples of $M$, $X$ and $K$ points. At all of these momenta the same trend is undoubtedly present, with the flux gap at zero field being broken apart and physical spin spectral weight being pushed down to zero energy as the GSL is entered. 

\begin{figure}[!t]  
\includegraphics[scale=0.4]{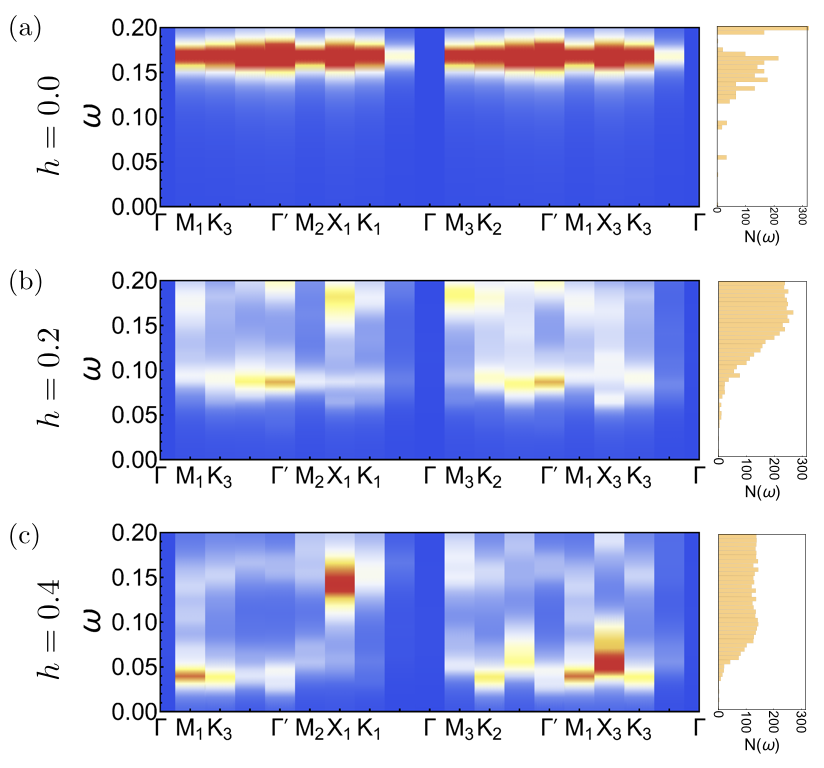}
\caption{Dynamical spin structure factor. (a) The zero-field KSL, (b) a point midway in the KSL phase 
		and (c) a point in the middle of the GSL 
		along a path through all high-symmetry points of the extended Brillouin zone for the cut $\theta=7.5^\circ$, the lower of the two dashed red lines in Fig.~\ref{fig:UPDs}(a) of the main text.}
\label{fig:SMDySF}
\end{figure}

However we note that, depending on the field direction, this trend may not be true for all momenta. In particular, if the field along a particular spin axis is negligible, the points in momentum space associated with the corresponding spin component will not be as strongly affected (as spin and momentum space are intrinsically linked together in the Kitaev model, for example the bond in real space associated with say the $x$-component of the spin translates into a particular direction in the BZ). As a result the flux gap will only fully collapse for those momenta along which the corresponding field magnitude is large. An example of this is shown in Fig.~\ref{fig:SMDySF} where we plot the dynamical spin structure factor in the middle of the GSL along the cut at $\theta=7.5^\circ$ away from the $[\bar{1}{1}0]$ axis. In this case the field along the $z$-axis, $h_z$, is much smaller compared to the other two components. This results in the flux gap remaining mostly intact at, say, $X_1$ and $K_1$ for example, while it closes at the remaining momenta. This raises the interesting possibility that the gauge field excitations may be anisotropic. In other words the gauge field propagator may be gapless for excitations with $\vec{k}$ along certain directions, and gapped for other directions, which could be captured by a $\vec{k}$-dependent mass. 

\section{Supplementary Note 4: Stability of the Fermi Surface}

Regarding the stability of the Fermi surface, though a normal Fermi liquid is unstable to an infinitesimal attractive interaction, coupling to a dynamical $U(1)$ gauge field stabilizes the Fermi surface up to a finite critical strength of the interaction \cite{Metlitski2015}. Further, we note that while there are several symmetry-allowed fermionic pairing terms, these will carry a non-trivial momentum dependence due to the shift of the Dirac cone away from high-symmetry momenta for generic field directions and as such will not contribute to the low-energy sector, as at least one of the momenta is far away from the Fermi surface. Note, however, that for magnetic fields applied along high-symmetry directions, some of these pairing terms might become relevant and thereby lead to a spin-Peierls instability akin to what has been discussed for Majorana Fermi surfaces \cite{Hermanns2015spin-peierls}.

\section{Supplementary Note 5: Exact Diagonalization Results in Context}

For those that may not necessarily be familiar with the energy scales typically encountered in exact diagonalization (ED) studies we would like to highlight the smallness of the gaps encountered in this work when compared to some known ED results. In particular we can compare to:
\begin{itemize}
\item The regular Heisenberg antiferromagnet on square, triangular and honeycomb lattices. In all cases the ground state is magnetically ordered with gapless Goldstone modes. Examples of spin gaps from ED include, $\Delta = 0.288$ for an $N=36$ site symmetric square cluster, $\Delta = 0.123$ for an $N=36$ site symmetric triangular cluster and $\Delta = 0.214$ for an $N=32$ site symmetric honeycomb cluster. Even the reported extrapolated gaps, taken from considering a range of system sizes, are much larger than those in the GSL (for all cluster sizes we have studied), being $\Delta_{\infty}=0.025$ for the square lattice (taken from 12 lattices ranging from $N=18-40$), $\Delta_{\infty}=0.129$ for the triangular lattice (taken from 12 lattices ranging from $N=24-36$) and $\Delta_{\infty}=0.050$ for the honeycomb lattice (taken from 14 lattices ranging from $N=6-38$). All values are taken from \cite{Richter2004} and references therein.  
 
\item The Kagome lattice Heisenberg antiferromagnet (KHAFM). There is an ongoing debate as to whether this is a gapless Dirac spin liquid or a gapped $\intg_2$ spin liquid. In either case the gap between the ground state and first excited state should go to zero for a torus geometry (for the Dirac SL because it is gapless and for the $\intg_2$ QSL because there should be a four-fold ground state degeneracy). A recent ED study has investigated $N=36$, $42$ and $48$ site clusters \cite{Laeuchli482016}. The encountered gaps are $\Delta=0.010$, $0.020$ and $0.021$, respectively. In other words, if we take the smallest ED gap for the KHAFM (the $N=36$ site cluster), then we can fit the lowest lying state from every single momentum sector within this gap for our $N=30$ and $32$ site clusters. If we take the ED gap from the largest system size studied, the $N=48$ site cluster, then for all of our clusters, $N=18-32$ sites, we can fit the lowest lying state from every single momentum sector within this gap.
 
\item Recent studies of Kalmeyer-Laughlin chiral spin liquids. Such phases have a two-fold degenerate ground state manifold (GSM) for a toroidal geometry which is exponentially split on finite sized systems, meaning the gap between the ground state and first excited state should scale to zero. As a particularly enlightening comparison we can compare our $N=32$ site honeycomb results with results for CSLs on $N=32$ site honeycomb lattices. Comparing to data from \cite{ShengHaldane2012} and \cite{HickeyPRL2016} they find gaps of $\Delta = 0.011$ and $\Delta = 0.035$ respectively. For our $N=32$ site results we can again fit at least $16$ states (one from each momentum sector) within these gaps. 
\end{itemize}

In summary, when compared to other ED studies for states, which we either know to be gapless or know/suspect to have a ground state degeneracy, the gaps encountered in our study are exceptionally small, not just for the first excited state but for the lowest lying state in every single momentum sector.


%

\end{document}